# Influence of 4-vinylbenzylation on the rheological and swelling properties of photo-activated collagen hydrogels

Giuseppe Tronci,[1,2] Colin A. Grant,[3] Neil H. Thomson,[2,4] Stephen J. Russell,[1] David J. Wood[2]

[1] Nonwovens Research Group, Centre for Technical Textiles, School of Design, University of Leeds, Leeds LS2 9JT, United Kingdom
[2] School of Dentistry, St. James's University Hospital, University of Leeds, Leeds LS9 7TF, United Kingdom
[3] Advanced Materials Engineering, Polymer IRC Labs, Faculty of Engineering & Informatics, University of Bradford, Bradford BD7 1DP, United Kingdom
[4] Molecular and Nanoscale Physics, School of Physics and Astronomy, University of Leeds, Leeds LS2 9JT, United Kingdom

## ABSTRACT

Covalent functionalisation of collagen has been shown to be a promising strategy to adjust the mechanical properties of highly swollen collagen hydrogels. At the same time, secondary interactions between for example, amino acidic terminations or introduced functional groups also play an important role and are often challenging to predict and control. To explore this challenge, 4-vinylbenzyl chloride (4VBC) and methacrylic anhydride (MA) were reacted with type I collagen, and the swelling and rheological properties of resulting photo-activated hydrogel systems investigated. 4VBC-based hydrogels showed significantly increased swelling ratio, in light of the lower degree of collagen functionalisation, with respect to methacrylated collagen networks, whilst rheological storage moduli were found to be comparable between the two systems. To explore the role of benzyl groups in the mechanical properties of the 4VBC-based collagen system, model chemical force microscopy (CFM) was carried out in aqueous environment with an aromatised probe against an aromatised gold-coated glass slide. A marked increase in adhesion force ($F$: 0.11±0.01 nN) was measured between aromatised samples, compared to the adhesion force observed between the non-modified probe and a glass substrate ($F$: 2.64±1.82 nN). These results suggest the formation of additional and reversible π-π stacking interactions in aromatic 4VBC-based networks and explain the remarkable rheological properties of this system in comparison to MA-based hydrogels.

## INTRODUCTION

Collagen is the main component of the extracellular matrix of tissues and has been widely applied to promote skin wound healing [1, 2]. However, despite its unique mechanical properties *in vivo*, collagen is mechanically unstable *ex vivo* in physiological conditions, mainly due to the breakdown of covalent crosslinks across collagen molecules. Blend formation with non-degradable, hydrophilic building blocks, e.g. polyethylene glycol [3], or cross-linking with potentially toxic compounds, e.g. dialdehydes [4], are usually employed to ensure mechanical integrity following material contact with biological fluids. Although these approaches can enhance the mechanical stability in biological environments, collagen organisation and bio-functionality can be irreversibly altered, whilst adjusted mechanical properties and bespoke material processability can only partially be accomplished [5]. To overcome these limitations, covalent functionalisation of collagen with photo-active compounds has recently been reported as a promising strategy yielding highly swollen and mechanically competent hydrogels suitable as wound dressings [6], whereby no secondary polymeric phase or direct cross-linking reaction is required. Similarly to the case of purely synthetic polymer networks, the macroscopic properties of resulting hydrogels can be controlled by the architecture of the covalent network, although secondary interactions are expected to play an important role e.g. between amino acidic collagen terminations or introduced functional groups as well as the protein conformation in the cross-linked state [7]. In order to gain insight into the role of covalently coupled photo-active moieties on the structure-property relationships of resulting collagen hydrogels, two photo-active collagen systems were compared, as obtained via reaction of type I collagen with either an aromatic monomer, i.e. 4-vinylbenzyl chloride (4VBC), or commonly-used methacrylate monomers, i.e. methacrylic anhydride (MA) and glycidyl methacrylate (GMA). Rheological and swelling properties were studied and discussed with regard to the chemical reactivity and structure of monomers, degree of functionalisation and collagen organisation.

**EXPERIMENTAL DETAILS**

Type I collagen isolated in-house from rat tail tendons was functionalised with either 4VBC or (G)MA as reported previously [6]. The degree of collagen functionalisation was quantified via 2,4,6-trinitrobenzene sulfonic acid (TNBS) colorimetric assay [8]. The swelling ratio ($SR$) of obtained hydrogels was quantified as ($m_s \cdot m_d^{-1}$), where $m_s$ and $m_d$ are the weights of freshly-synthesised, water-equilibrated hydrogels and dry networks, respectively. Rheological measurements were carried out via a AR1500ex Rheometer (TA Instruments, Crawley, UK) equipped with a 60 mm plate and a solvent trap. Frequency sweeps were carried out with a 3 mm hydrogel-plate gap and a 0.5% strain amplitude (according to the linear viscoelastic region), with three replicas for each hydrogel. CFM measurements were performed with a MFP-3D AFM (Asylum Research, Santa Barbara, CA, USA) and an AFM cantilever (k ~ 0.12 N·m$^{-1}$) bearing a Ø 10 $\mu$m gold-coated borosilicate glass probe (PT.BORO.(AU.)SN10, Novascan Technologies, Inc.). Calibration of the detector sensitivity and cantilever spring constant was carried out using the thermal method [9]. AFM probe and gold-coated microscopy slides (AU.0100.ALSI, Platypus Technologies LLC) were thiolated via incubation in a 3 mM 2-phenylethanethiol (2PET) solution in ethanol (25 °C, overnight), as previously reported [10]. Reacted samples were rinsed thoroughly with distilled water to remove any unreacted moiety. Force volume measurements were made in organized arrays (30 × 30 $\mu$m) of indentations at a piezo velocity of 2 $\mu$m·s$^{-1}$ with 5 nN force.
In the following, samples are coded as 'CRT-XXXYY$^*$', whereby 'CRT' indicates that the collagen used was extracted from rat tails; 'XXX' identifies the type of monomer introduced, either 4VBC or (G)MA; YY represents the monomer excess used in the functionalisation reaction with respect to collagen amino groups; '$^*$' is used to identify an hydrogel sample obtained by the photo-activation of the corresponding functionalised collagen precursor.

**DISCUSSION**

<u>**Rheological properties of 4-vinylbenzylated and methacrylated collagen hydrogels**</u>

Frequency sweep rheograms were obtained for both hydrogels systems in the linear viscoelastic region. As displayed in Figure 1, storage modulus ($G'$) was measured to be about one order of magnitude higher than the loss modulus ($G''$) with minimal dependency on frequency in the range $10^{-2} - 10^1$ Hz regardless of the network architecture. Both moduli were observed to increase markedly at higher frequencies ($10^1 - 10^2$ Hz). The rheological behaviour observed by the frequency sweeps is comparable to that described in other biomimetic hydrogels [11, 12] and confirms the formation of a covalent network following UV irradiation of both functionalised collagen solutions. This was also supported by the fact that hydrogels incubated overnight in 10 mM HCl solution (under which pH conditions collagen is completely soluble) displayed no change in dry weight with respect to freshly-synthesised samples. The fact that the storage modulus profile showed an almost frequency-independent behaviour at $10^{-2} - 10^1$ Hz with higher $G'$ values compared to $G''$ profile indicates a dominant elastic-like response of the collagen hydrogel. Furthermore, the profile of $tan\delta$ described by sample CRT-4VBC25* revealed lower values compared to sample CRT-MA25*, suggesting an increased elastic behaviour in the former system. This correlates well with the introduction of the stiff, aromatic crosslinking segment in the benzylated network, potentially resulting in additional π-π physical crosslinks, as discussed in the following sections.

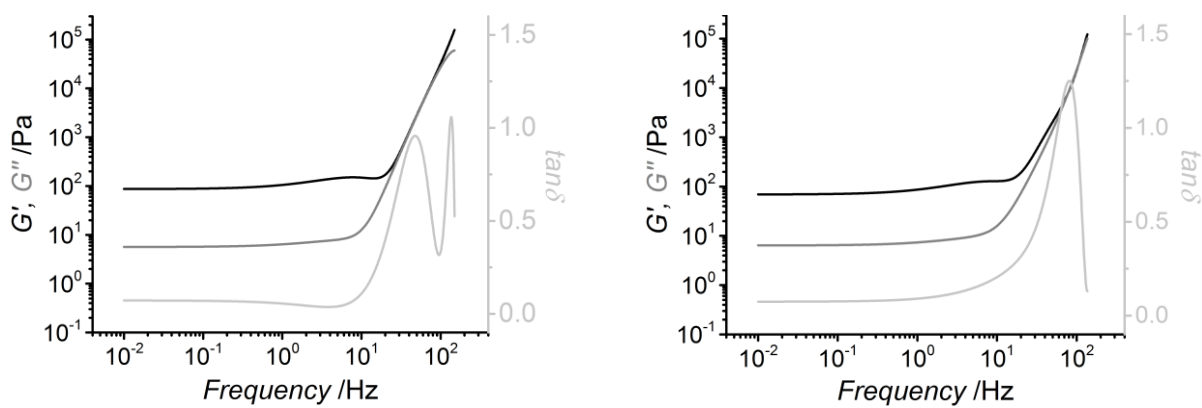

**Figure 1.** Storage modulus ($G'$, black), loss modulus ($G''$, grey) and $tan\delta$ (light grey) obtained via averaging and fitting (5$^{th}$ degree polynomial function, $R^2 > 0.99$) three frequency sweeps of hydrogels CRT-4VBC25* (left) and CRT-MA25* (right), respectively.

## **Structure-property relationships**

Once the formation of a complete covalent network was confirmed, the attention moved to understanding the structure-property relationships of both hydrogel systems. The degree of collagen functionalisation ($F$) in either 4-vinylbenzylated or methacrylated collagen was determined, as a key molecular parameter strictly related to the cross-link density of the photo-activated collagen networks. Besides the information at the molecular level, the rheological storage modulus ($G'$) and the weight ratio between freshly-synthesised, water-equilibrated hydrogels and corresponding dry networks were also measured in order to characterise the macroscopic behaviour of obtained materials. Reaction of collagen with MA resulted in nearly quantitative conversion of primary amino groups to methacrylate functions ($F_{MA} \sim 90$ mol.-%), whilst nearly 30 mol.-% lysines were functionalised with 4VBC in the same reaction conditions (Figure 2). Although chlorine is a good leaving group, the higher collagen functionalisation observed following reaction with MA compared to the reaction with 4VBC is attributed to the different mechanism of the coupling reactions, i.e. nucleophilic addition/elimination and nucleophilic substitution, respectively. This together with considerations around the electrophilicity of MA carbonyl group and leaving group capability of the methacrylate ion following nucleophilic attack of collagen amino terminations may account for these observations.

The difference in covalent functionalisation of collagen precursors at the molecular level appeared to be reflected in the swelling properties of respective collagen hydrogels, with 4VBC-based networks displaying more than twice the water content ($m_s \cdot m_d^{-1}$: 266 ± 18 wt.-%) of the methacrylated variant ($m_s \cdot m_d^{-1}$: 127 ± 6 wt.-%) (Figure 2). Similar trends in swelling properties were already observed when comparing samples of CRT-4VBC25[*] with glycidylmethacrylated collagen networks obtained with comparable degree of functionalisation [6], whereby the swelling ratio was still found to be increased in benzylated rather than methacrylated systems. This observation is partially in contrast with the high water content displayed by methacrylated, branched polyethyleneimine (PEI) hydrogels, although in that case the swelling behaviour was correlated with the consumption of amino groups of PEI [13], given that methacrylated functions were simply grafted to the polymer backbone and kept unsaturated.

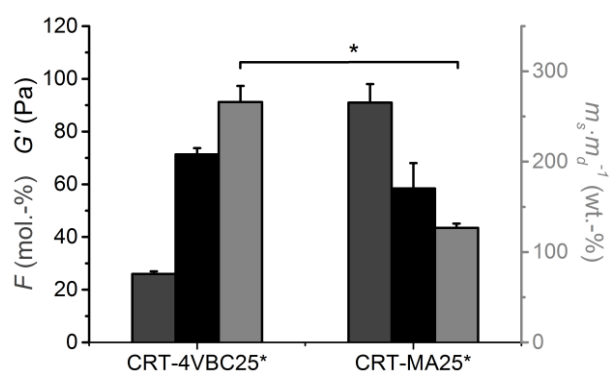

**Figure 2.** Degree of functionalisation (*F*, dark grey columns), storage modulus (*G'* (f=0.01 Hz), black columns) and swelling ratio (*SR*, light grey columns) of collagen hydrogels obtained via either 4-vinylbenzylation (CRT-4VBC25[*]) or methacrylation (CRT-MA25[*]) of collagen lysines. '*' indicates that the means of the corresponding samples are significantly different (at the 0.05 level, Bonferroni test).

Besides determination of the weight ratio ($m_s \cdot m_d^{-1}$), iodomethane contact angle measurements were carried out in order to explore whether the type of collagen functionalisation may influence the wettability of resulting hydrogels, thereby affecting the hydrogel water content. Iodomethane was selected as a standard liquid for contact angle quantification and non-solvent for collagen, allowing for the quantification of static contact angles. Interestingly, similar contact angles were measured in samples CRT-4VBC25[*] and CRT-GMA15[*] in both dry and wet states (Figure 3), with the latter samples being obtained with comparable degree of glycidyl methacrylation with respect to benzylation [5]. These results therefore confirm that the introduction of either methacrylate or benzyl moieties does not induce a detectable variation in wettability in obtained collagen materials, which can be explained by the partial hydrogen bonding capability of both methacrylate moieties [13] and aromatic rings [14] with water. Consequently, the observed swelling behaviour is likely to be ruled by the network architecture of the collagen hydrogel, as described by the degree of functionalisation and cross-link density.

Other than the larger water content displayed by methacrylated with respect to benzylated networks, comparable *G'* values were unexpectedly observed between the two systems (Figure 2). In principle, water acts as plasticiser in biopolymers [15], so that specifically lowered storage moduli should be expected in materials displaying increased swelling ratio. At the same time, considerations on the backbone stiffness of introduced moieties are likely to account for this observation. Indeed, functionalisation of collagen with 4VBC leads to the formation of a crosslinking segment consisting of an intermediate 5-carbon chain bearing two benzyl terminations. Such a cross-linking segment is expected to be stiffer and longer than the aliphatic segment resulting from the cross-linking of two methacrylated collagen molecules.

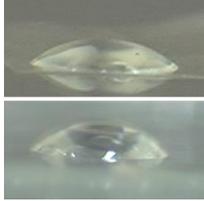

| Sample ID | CRT-4VBC25[*] | | CRT-GMA15[*] | |
|---|---|---|---|---|
| State | Dry | Wet | Dry | Wet |
| $\theta$ | 41 ± 6° | 42 ± 7° | 40 ± 5° | 45 ± 11° |

**Figure 3.** Left: Contact angle ($\theta$) measured on collagen networks in both dry and water-equilibrated state following contact with a 50 µl iodomethane drop. Three measurements of both left and right contact angle were taken and values expressed as average ± standard deviation (n=6). Right: digital photographs of an iodomethane drop on either dry (top) or water-swollen (bottom) sample CRT-4VBC25[*].

Further to the molecular stiffness of the 4VBC-based crosslinking segment, 4VBC aromatic rings are also expected to mediate π-π stacking interactions following mechanical stimulus, thereby providing additional and reversible physical cross-links in the covalent collagen network.

**Chemical force microscopy model experiments**

In order to explore the capability of aromatic rings to mediate detectable π-π stacking interactions in the 4VBC-based collagen network, model CFM force-volume measurements were carried out in aqueous environment. Thiolation of a gold-coated AFM probe was performed as a convenient approach to introduce phenethyl terminations on the probe surface and the same approach was undertaken to graft a gold-coated microscopy slide with benzene groups [10, 16].

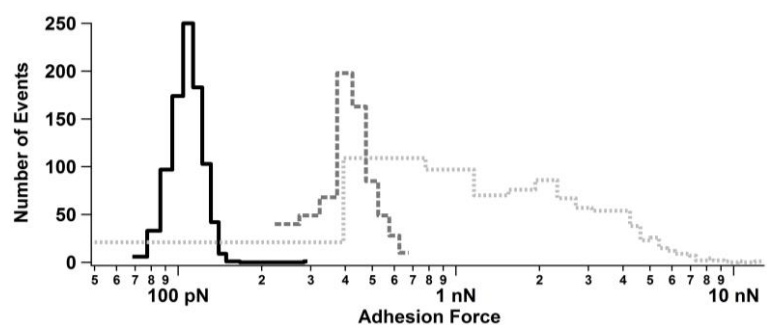

**Figure 4.** Distributions of adhesion force measured between a non-benzylated AFM probe and a gold-coated microscopy slide (black solid line, *F*: 0.11±0.01 nN), a benzylated AFM probe and a gold-coated microscopy slide (grey dashed line, *F*: 0.42±0.01 nN), and a benzylated AFM probe and a benzylated microscopy slide (light grey dotted line, *F*: 2.64±1.82 nN).

Figure 4 displays the distribution of adhesion forces between the three experimental sets, i.e. (1) benzylated AFM probe against a benzylated substrate, (2) benzylated AFM probe against a non-benzylated substrate and (3) non-benzylated AFM probe against a non-benzylated substrate. More than 20-fold increase in averaged adhesion force (0.11±0.01 → 2.64±1.82 nN) was quantified in the former case compared to the case of non-benzylated samples, confirming the establishment of significant aromatic interactions between the scanning probe and the model substrate under water. Similar adhesion forces have been observed between carboxylated AFM probe and carboxylated substrate, although these measurements were carried out in ethanol, whereby solvent-mediated hydrogen bonding is expected to be lower with respect to the case of distilled water [10]. Preliminary CFM experiments on the collagen hydrogels indicate that the benzylated probe force measurements will reveal additional information on their wettability and surface structure.

## CONCLUSIONS

Two collagen hydrogel systems obtained via either benzylation or methacrylation of collagen lysines were investigated in order explore the influence of photo-active moieties on the material structure-property relationships. Rheological properties were quantified via frequency sweeps, whilst the swelling behaviour was investigated via gravimetric as well as contact angle measurements. Model CFM experiments were carried out to probe the formation of aromatic interactions in benzylated collagen networks. The introduction of 4VBC moieties was found to strongly affect the mechanical properties of selected collagen networks, in light of the backbone stiffness of the cross-linking segments and potential π-π stacking interactions between introduced aromatic rings. The degree of collagen functionalisation was found to correlate with the variation in water content between methacrylated and benzylated networks.

## ACKNOWLEDGMENTS

The authors gratefully acknowledge financial support from the Clothworkers' Centre for Textile Materials Innovation for Healthcare.